\documentstyle[epsfig,12pt,a4p]{article}

\newcommand{\kkkk}{\mbox{$K^+K^-K^+K^-$ }}
\newcommand{\kpikpi}{\mbox{$K^+K^-\pi^+\pi^-$ }}
\newcommand{\pipipipi}{\mbox{$\pi^+\pi^-\pi^+\pi^-$ }}

\begin{document}
\begin{titlepage}
\def\footnoterule{\hrule width 1.0\columnwidth}
\begin{tabbing}
put this on the right hand corner using tabbing so it looks
 and neat and in \= \kill
\> {13 May 1998}
\end{tabbing}
\bigskip
\bigskip
\begin{center}{\Large  {\bf A study of the centrally produced
$\phi\phi$ system
in pp interactions at 450 GeV/c}
}\end{center}
\bigskip
\bigskip
\begin{center}{        The WA102 Collaboration
}\end{center}\bigskip
\begin{center}{
D.\thinspace Barberis$^{  5}$,
W.\thinspace Beusch$^{   5}$,
F.G.\thinspace Binon$^{   7}$,
A.M.\thinspace Blick$^{   6}$,
F.E.\thinspace Close$^{  4,5}$,
K.M.\thinspace Danielsen$^{ 12}$,
A.V.\thinspace Dolgopolov$^{  6}$,
S.V.\thinspace Donskov$^{  6}$,
B.C.\thinspace Earl$^{  4}$,
D.\thinspace Evans$^{  4}$,
B.R.\thinspace French$^{  5}$,
C.\thinspace Goodwin$^{  4}$,
T.\thinspace Hino$^{ 13}$,
S.\thinspace Inaba$^{   9}$,
A.V.\thinspace Inyakin$^{  6}$,
T.\thinspace Ishida$^{   9}$,
A.\thinspace Jacholkowski$^{   5}$,
T.\thinspace Jacobsen$^{  12}$,
G.T\thinspace Jones$^{  4}$,
G.V.\thinspace Khaustov$^{  6}$,
T.\thinspace Kinashi$^{  14}$,
J.B.\thinspace Kinson$^{   4}$,
A.\thinspace Kirk$^{   4}$,
W.\thinspace Klempt$^{  5}$,
V.\thinspace Kolosov$^{  6}$,
A.A.\thinspace Kondashov$^{  6}$,
A.A.\thinspace Lednev$^{  6}$,
V.\thinspace Lenti$^{  5}$,
S.\thinspace Maljukov$^{   8}$,
P.\thinspace Martinengo$^{   5}$,
I.\thinspace Minashvili$^{   8}$,
K.\thinspace Myklebost$^{   3}$,
T.\thinspace Nakagawa$^{  13}$,
K.L.\thinspace Norman$^{   4}$,
J.M.\thinspace Olsen$^{   3}$,
J.P.\thinspace Peigneux$^{  1}$,
S.A.\thinspace Polovnikov$^{  6}$,
V.A.\thinspace Polyakov$^{  6}$,
V.\thinspace Romanovsky$^{   8}$,
H.\thinspace Rotscheidt$^{   5}$,
V.\thinspace Rumyantsev$^{   8}$,
N.\thinspace Russakovich$^{   8}$,
V.D.\thinspace Samoylenko$^{  6}$,
A.\thinspace Semenov$^{   8}$,
M.\thinspace Sen\'{e}$^{   5}$,
R.\thinspace Sen\'{e}$^{   5}$,
P.M.\thinspace Shagin$^{  6}$,
H.\thinspace Shimizu$^{ 14}$,
A.V.\thinspace Singovsky$^{ 1,6}$,
A.\thinspace Sobol$^{   6}$,
A.\thinspace Solovjev$^{   8}$,
M.\thinspace Stassinaki$^{   2}$,
J.P.\thinspace Stroot$^{  7}$,
V.P.\thinspace Sugonyaev$^{  6}$,
K.\thinspace Takamatsu$^{ 10}$,
G.\thinspace Tchlatchidze$^{   8}$,
T.\thinspace Tsuru$^{   9}$,
M.\thinspace Venables$^{  4}$,
O.\thinspace Villalobos Baillie$^{   4}$,
M.F.\thinspace Votruba$^{   4}$,
Y.\thinspace Yasu$^{   9}$.
}\end{center}

\begin{center}{\bf {{\bf Abstract}}}\end{center}

{
The reaction $pp\rightarrow p_fp_s(K^+K^-K^+K^-)$ in which the
\kkkk
system is centrally produced has been studied at 450 GeV/c.
$\phi\phi$ production has been found to dominate this reaction and
is compatible with being produced by double Pomeron exchange.
An angular analysis of the $\phi\phi$ system favours $J^{PC} = 2^{++}$
and its $dP_T$ dependence is similar to that observed for glueball candidates.
}
\bigskip
\bigskip
\bigskip
\bigskip\begin{center}{{Submitted to Physics Letters}}
\end{center}
\bigskip
\bigskip
\begin{tabbing}
aba \=   \kill
$^1$ \> \small
LAPP-IN2P3, Annecy, France. \\
$^2$ \> \small
Athens University, Physics Department, Athens, Greece. \\
$^3$ \> \small
Bergen University, Bergen, Norway. \\
$^4$ \> \small
School of Physics and Astronomy, University of Birmingham, Birmingham, U.K. \\
$^5$ \> \small
CERN - European Organization for Nuclear Research, Geneva, Switzerland. \\
$^6$ \> \small
IHEP, Protvino, Russia. \\
$^7$ \> \small
IISN, Belgium. \\
$^8$ \> \small
JINR, Dubna, Russia. \\
$^9$ \> \small
High Energy Accelerator Research Organization (KEK), Tsukuba, Ibaraki 305,
Japan. \\
$^{10}$ \> \small
Faculty of Engineering, Miyazaki University, Miyazaki, Japan. \\
$^{11}$ \> \small
RCNP, Osaka University, Osaka, Japan. \\
$^{12}$ \> \small
Oslo University, Oslo, Norway. \\
$^{13}$ \> \small
Faculty of Science, Tohoku University, Aoba-ku, Sendai 980, Japan. \\
$^{14}$ \> \small
Faculty of Science, Yamagata University, Yamagata 990, Japan. \\
\end{tabbing}
\end{titlepage}
\setcounter{page}{2}
\bigskip
\par
\par
Experiment WA102 is designed to study exclusive final states formed in
the reaction
\begin{equation}
pp \rightarrow p_{f} (X^0) p_{s}
\label{eq:a}
\end{equation}
at 450 GeV/c.
The subscripts f and s indicate the
fastest and slowest particles in the laboratory respectively
and $X^0$ represents the central
system that is presumed to be produced by double exchange processes.
The experiment
has been performed using the CERN Omega Spectrometer
the layout of which is
described in ref.~\cite{WADPT}
In previous analyses of other channels it has been observed that
when the centrally produced system has been analysed
as a function of the parameter $dP_T$, which is the difference
in the transverse momentum vectors of the two exchange
particles~\cite{WADPT,closeak},
all the undisputed $q \overline q$ states are suppressed at small
$dP_T$ in contrast to glueball candidates.
In addition, it has recently been suggested~\cite{fec,jmf}
that this effect could be due to the fact that
the production mechanism is through the fusion of two vector particles.
It is therefore interesting to make a study of the $dP_T$ dependence of
systems decaying to two vectors. One such system is the $\phi \phi$
final state which
has the added interest that the observation of
glueball candidates has been claimed in this channel~\cite{LINDE}.
\par
This paper presents a study of the $\phi\phi$ system decaying to
\kkkk produced in central
pp interactions and represents a factor ten increase over
previously published data samples~\cite{WA76,WA76P}.
Events corresponding to the reaction
\begin{equation}
pp \rightarrow p_{f} (K^+K^-K^+K^-) p_{s}
\label{eq:b}
\end{equation}
have been isolated from the sample of events having six
outgoing
charged tracks
by first imposing the following cuts on the components of the
missing momentum:
$|$missing~$P_{x}| <  14.0$ GeV/c,
$|$missing~$P_{y}| <  0.12$ GeV/c and
$|$missing~$P_{z}| <  0.08$ GeV/c,
where the x axis is along the beam
direction.
A correlation between
pulse-height and momentum
obtained from a system of
scintillation counters was used to ensure that the slow
particle was a proton.
\par
In order to select the \kkkk system, information
from the {\v C}erenkov counter was used.
Two data samples were selected: Sample A
in which three of the centrally
produced particles were identified as ambiguous kaons/protons
and sample B
in which one
of the centrally produced particles was identified as
an ambiguous kaon/proton
and none of the other particles was positively
identified as a pion.
\par
The method of Ehrlich et al.~\cite{EHRLICH} ,
modified for four tracks, has been used to compute the mass
squared of the four central particles (assumed to be equal).
The resulting distribution for sample A is shown in
fig.~\ref{fi:1}a) where a clear peak can be seen at
the kaon mass squared. This distribution has been fitted with
three Gaussians to represent the contributions from the
\pipipipi, \kpikpi and \kkkk channels.
{}From this fit the number of \kkkk events in sample A is estimated to
be 305~$\pm$~30.
A cut on the mass squared of
$0.16 \leq M^2 \leq 0.36$~$GeV^2$
has been used to select a sample of \kkkk events.
The mass of one $K^+K^-$ pair is plotted against the mass of the other
$K^+K^-$ pair in
fig.~\ref{fi:1}c) and shows a strong $\phi\phi$ signal.
\par
As can be seen from fig.~\ref{fi:1}b)
the Ehrlich mass distribution for sample B is dominated by a
peak at the pion mass squared. However,
if the same cut used for sample A is applied
then, as can be seen from fig.~\ref{fi:1}d),
the $\phi\phi$ channel can clearly be selected.
\par
The four possible $K^+K^-$ mass combinations are plotted
in fig.~\ref{fi:2}a) and b) for samples A and B respectively.  A clear $\phi$
signal can be seen.
A fit has been performed to these spectra
where the $\phi$ is described by
a spin 1 relativistic Breit-Wigner convoluted with a Gaussian to represent
the experimental resolution,
and a background of the form
$(m -m_{th})^a exp(-bm-cm^2)$
where $m$ is the $K^+K^-$ mass, $m_{th}$ is the threshold mass and
a,b and c are fit parameters.
The width of the Breit-Wigner has been fixed to the PDG value of
4.43~MeV~\cite{PDG96}.
The result of the fit to the two spectra
are compatible and give $m(\phi)$~=~1019.5~$\pm$~0.4~MeV
and $\sigma(Gaussian)$~=~3.3~$\pm$~0.5~MeV.
By selecting one $K^+K^-$ mass to lie within a band around the $\phi$ mass
(from 1.01 - 1.03~GeV) and plotting the effective mass of the
other pair, the spectra of fig.~\ref{fi:2}c) and d) were produced
for samples A and B respectively. The large $\phi$ signal
with little background confirms the presence of a strong $\phi\phi$ signal.
The background beneath the $\phi$ signal is similar in both samples.
\par
The number of events in nine regions around the $\phi\phi$ position in the
$K^+K^-$ scatter plot are shown in fig.~\ref{fi:2}e) and f) for samples A and B
respectively; no event has more than one
entry in this region. From these
numbers and applying a correction for the tails of the $\phi$,
the total number of $\phi\phi$ events is found to be 110~$\pm$~18 (sample A)
and 409~$\pm$~33 events (sample B).
\par
In order to compare the production rates for $\phi K^+K^-$  and $\phi \phi$,
the number of $\phi K^+ K^-$ events has been estimated. This was done
by subtracting twice the number of $\phi\phi$ events from the total
number of $\phi$s observed in fig.~\ref{fi:2}a) and b), and this gives
187~$\pm$~47 $\phi K^+ K^-$ events for sample A and
600~$\pm$~84 $\phi K^+ K^-$ events for sample B.
A possible source of contamination could come from $\phi \pi^+ \pi^-$
events where a $\pi$ was misidentified as a K.
However, it has been shown previously~\cite{oldkstkst} that the
$\phi \pi^+ \pi^-$ channel is very weak and after the selections applied
in this analysis its contribution to this channel is found to be negligible.
\par
The geometrical acceptance has been found to be similar for $\phi \phi$
and $\phi K^+K^-$ production.
After correcting for unseen decay modes of
the $\phi$, the ratio of cross sections is estimated to be
$\sigma(\phi K^+K^-)/\sigma(\phi \phi)$~=~0.83~$\pm$~0.21 (sample A)
and 0.72~$\pm$~0.07 (sample B).
For sample A we can also calculate the
number of \kkkk events that do not include a $\phi$ using the fit
to fig.~\ref{fi:1}a) which gives 7~$\pm$~50 \kkkk events.
The results obtained for samples A and B are similar and since
the statistics are greater in sample B for the rest of the results
presented in this paper we shall use sample B only.
This dominance of the $\phi \phi$ channel over the $\phi K^+K^-$ and
\kkkk channels is similar to what has been observed by the JETSET
experiment~\cite{JETSET}.
\par
The $\phi\phi$ final state is selected by requiring that both $K^+K^-$ mass
combinations fall within 1.01~$\leq$~$m(K^+K^-)$~$\leq$~1.03~GeV.
The $\phi \phi$
effective mass spectrum is shown in fig.~\ref{fi:3}a) and as can be seen
shows a broad distribution with a maximum around 2.35~GeV.
\par
The angular distributions of the \kkkk system can be used to determine
the spin-parity of the intermediate $\phi \phi$ state using
a method formulated by Chang and Nelson~\cite{CHANG}
and Trueman~\cite{TRUEMAN}.
Three angles have to be considered: the azimuthal angle $\chi$, between
the two $\phi$ decay planes and the two polar angles $\theta_1$ and $\theta_2$
of the $K^+$s in their respective $\phi$ rest frames
relative to the $\phi$ momenta in the $\phi \phi$ rest frame.
\par
For a $\phi \phi$ sample of unique spin-parity and free of
background the distribution of $\chi$ takes the form
$dN/d \chi = 1+ \beta cos(2\chi)$
where $\beta$ is a constant which depends only on the spin-parity
of the $\phi \phi$ system and is independent of its polarisation.
Similarly
$dN/dcos \theta = 1 + (\zeta/2)(3cos^2\theta - 1)$.
Values of $\beta$ and $\zeta$ for different spin-parity states are
given in table~\ref{ta:1}~\cite{EIGEN}.
Fig.~\ref{fi:3}b) and c) shows the $\chi$ and $cos \theta$ distributions.
A chi-squared fit has been performed to these spectra using the
distributions expected for a single
state with a given value of $J^{PC}$.
The results of the fit are given in table~\ref{ta:1} and as can be seen
the lowest chi-squared is for a fit using
$J^P$~=~$2^+$ (L=0, S=2) with $\beta$=1/15 and $\zeta$=0.
As can be seen the $J^{PC}$~=~$0^{++}$ and $0^{-+}$
hypotheses can be clearly ruled out.
A free fit to the distributions has been performed
and gives $\beta$~=~0.0~$\pm$~0.1 and $\zeta$~=~0.1~$\pm$~0.1.
The spin analysis has been repeated in three slices in mass of the $\phi \phi$
system. The results found are similar to that for the total sample.
\par
The Feynman $x_F$ distributions for the slow particle,
the $\phi\phi$ system and the fast particle
are shown in fig.~\ref{fi:4}a).
As can be seen the $\phi\phi$ system lies within $|x_F| \leq 0.2$.
After correcting for
geometrical acceptances, detector efficiencies
and losses due to cuts
and charged kaon decay,
the cross-section for
$\phi \phi$ production at $\sqrt s$~=~29.1~GeV in the $x_F$ interval
$|x_F| \leq 0.2$ is $\sigma(\phi\phi)$~=~41.0~$\pm$~3.7~nb.
This can be compared with the cross-sections found in the same interval
at $\sqrt s$~=~12.7~\cite{WA76} and 23.8~GeV~\cite{WA76P} of
42~$\pm$~9~nb and 36~$\pm$~6~nb respectively.
This effectively constant cross-section as a function of energy
is consistent with the $\phi \phi$ system being produced by a
Double Pomeron Exchange (DPE) mechanism.
\par
A study of the $\phi \phi$ system as a function of the parameter
$dP_T$, which is the difference in the transverse momentum vectors
of the two exchanged
particles~\cite{WADPT,closeak}, has been performed.
The acceptance corrected $dP_T$ dependence of the $\phi\phi $ system is shown
in fig.~\ref{fi:4}b).
The fraction of $\phi \phi$
production has been calculated for
$dP_T$$\leq$0.2 GeV, 0.2$\leq$$dP_T$$\leq$0.5 GeV and $dP_T$$\geq$0.5 GeV and
gives
0.22~$\pm$~0.05, 0.51~$\pm$~0.04 and 0.27~$\pm$~0.04 respectively.
This results in a ratio of production at small $dP_T$ to large $dP_T$ of
0.81~$\pm$~0.22.
This ratio is much higher than what has been observed for
the undisputed $q \overline q$ states which typically have a value for
this ratio
of 0.1~\cite{memoriam}.
\par
Fig.~\ref{fi:4}c) shows the four momentum transferred from
one of the proton vertices
for the $\phi \phi$ system.
The distribution has been fitted with a single exponential
of the form $exp(-b |t|)$ and yields $b$~=~6.3~$\pm$~0.6~GeV
which is consistent
with what is expected from DPE~\cite{dpet}.
The azimuthal angle ($\phi$) between the $p_T$
vectors of the two protons
is shown in
fig.~\ref{fi:4}d).
This distribution differs significantly from that observed
for the $\pi^0$, $\eta$, $\eta^\prime$~\cite{0mpap} and
$\omega$~\cite{3pipap}.
\par
In conclusion, a study of the reaction $pp \rightarrow p_fp_s(K^+K^-K^+K^-)$
shows that the production of
$\phi \phi $ is the dominant process and that there is effectively
no \kkkk production not involving one or more $\phi$ mesons.
The $\phi \phi$
mass spectrum shows a broad distribution with a maximum around 2.35~GeV and
an angular analysis shows that it is compatible with having
$J^{PC}$~=~$2^{++}$.
The behaviour of the cross-section as a function of centre of mass
energy and the four momentum transferred dependence are compatible with what
would
be expected if the $\phi \phi $ system was produced via double Pomeron exchange
which has been predicted to be a source of glueballs~\cite{closerev}.
In addition, the $dP_T$ behaviour is different to that observed for all
undisputed
$q \overline q $ states.
\par
\begin{center}
{\bf Acknowledgements}
\end{center}
\par
This work is supported, in part, by grants from
the British Particle Physics and Astronomy Research Council,
the British Royal Society,
the Ministry of Education, Science, Sports and Culture of Japan
(grants no. 04044159 and 07044098)
and
the Russian Foundation for Basic Research (grant 96-15-96633).

\newpage
\newpage
\begin{table}[h]
\caption{The $\beta$, $\zeta$ and chi-squared values for different spins of the
$\phi \phi$ system}
\label{ta:1}
\vspace{1in}
\begin{center}
\begin{tabular}{|c|c|c|c|c|c|} \hline
 & & & & & \\
$J^{P}$ & L & S & $\beta$ & $\zeta$ & chi-squared \\
& & & & &\\ \hline
 & & & & & \\
$0^+$  & 0 &  0 &  2/3 & 0 & 124 \\
$0^+$  & 2 &  2 &  1/3 & 1 & 202 \\
$0^-$  & 1 &  1 &  -1 & -1 & 356 \\
$1^-$  & 1 &  1 &  0 & 1/2 & 69 \\
$1^+$  & 2 &  2 &  0 & 1/2 & 69 \\
$2^+$  & 0 &  2 &  1/15 & 0 & 36 \\
$2^+$  & 2 &  0 &  2/3 & 0 & 124 \\
$2^+$  & 2 &  2 &  2/21 & 3/14 & 44 \\
$2^-$  & 1 &  1 &  -2/5 & -1/10 & 61 \\
$2^-$  & 3 &  1 &  -31/5 & -2/3 & 162 \\
 & & & & & \\ \hline
\end{tabular}
\end{center}
\end{table}
\newpage
{ \large \bf Figures \rm}
\begin{figure}[h]
\caption{The Ehrlich mass squared distribution for a) sample A and b) sample B.
The lego Plot of one $K^+K^-$ mass against the other (two entries per event)
for c) sample A and d) sample B.
}
\label{fi:1}
\end{figure}
\begin{figure}[h]
\caption{
a) and b) The $K^+K^-$ mass spectrum (four combinations per event),
c) and d) the $K^+K^-$ effective mass of one $K^+K^-$ combination after
selecting the
other to lie in the $\phi$ mass band and
e) and f) a scatter table of one $K^+K^-$ mass against the other in the
$\phi \phi$ region.
a), c) and e) for sample A
b), d) and f) for sample B.
}
\label{fi:2}
\end{figure}
\begin{figure}[h]
\caption{
a) The $\phi \phi$ effective mass spectrum,
b) and c) the $\chi$ and $cos \theta$ distributions
(the superimposed curve represents the $2^+$ (L=0, S=2) wave).
}
\label{fi:3}
\end{figure}
\begin{figure}[h]
\caption{
a) The $x_F$ distribution for the slow particle, the $\phi \phi $ system and
the fast particle. b) The $dP_T$ spectrum,
c) the four momentum transfer squared ($|t|$) from one of the proton vertices
and
d) the azimuthal angle ($\phi$) between the two outgoing protons
for the $\phi \phi$ system.
}
\label{fi:4}
\end{figure}
\newpage
\begin{center}
\epsfig{figure=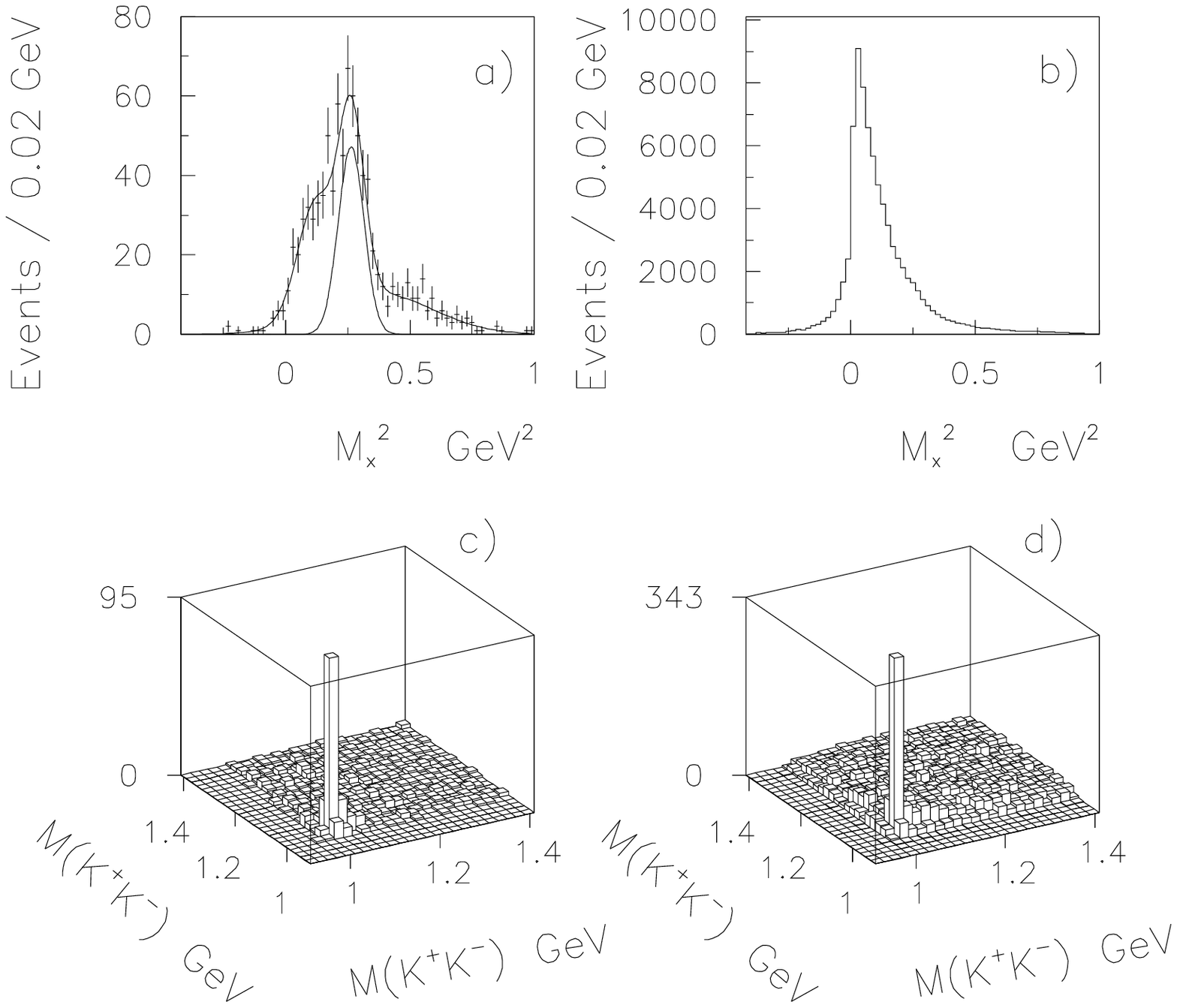,height=22cm,width=17cm}
\end{center}
\begin{center} {Figure 1} \end{center}
\newpage
\begin{center}
\epsfig{figure=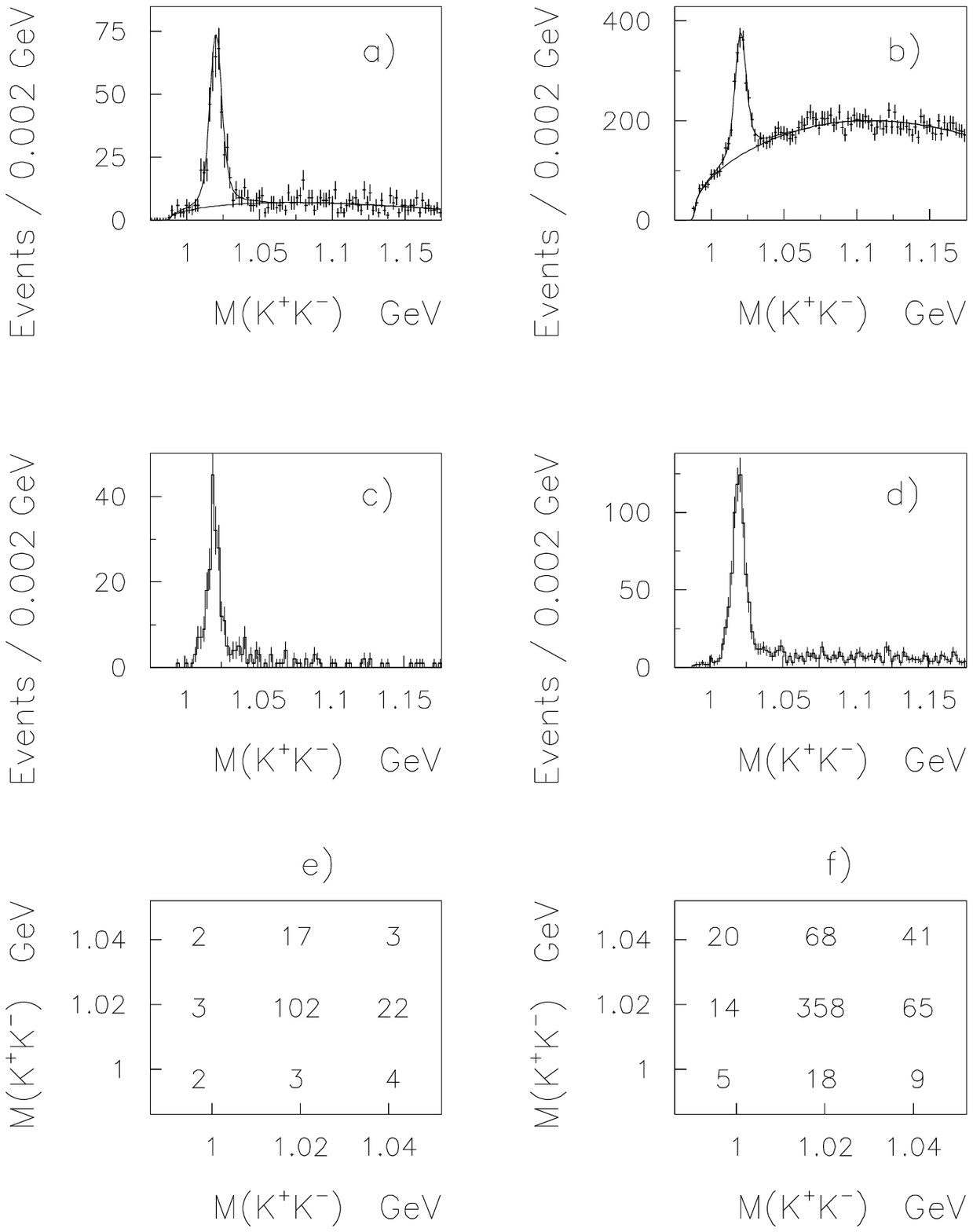,height=22cm,width=17cm}
\end{center}
\begin{center} {Figure 2} \end{center}
\newpage
\begin{center}
\epsfig{figure=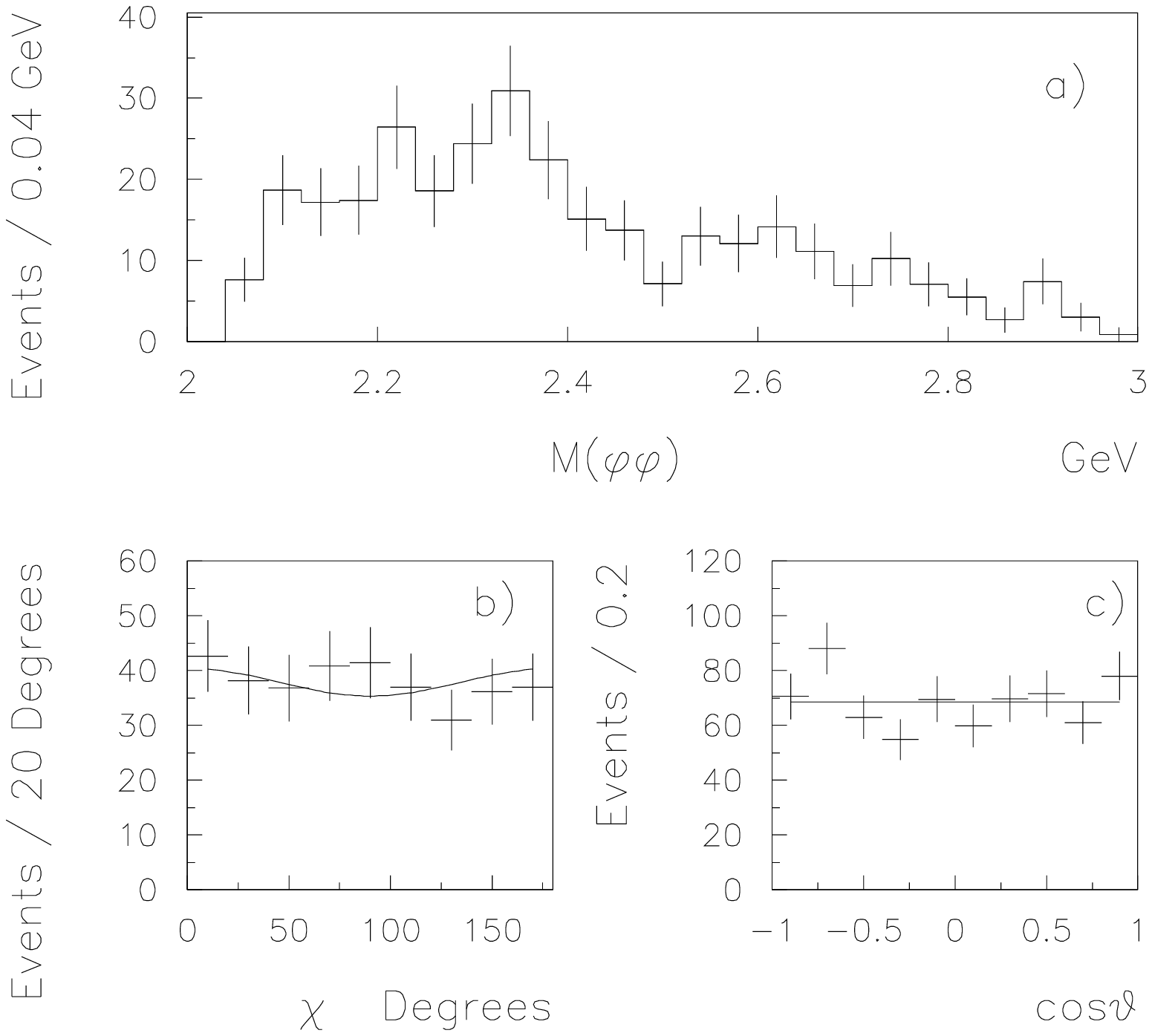,height=22cm,width=17cm}
\end{center}
\begin{center} {Figure 3} \end{center}
\newpage
\begin{center}
\epsfig{figure=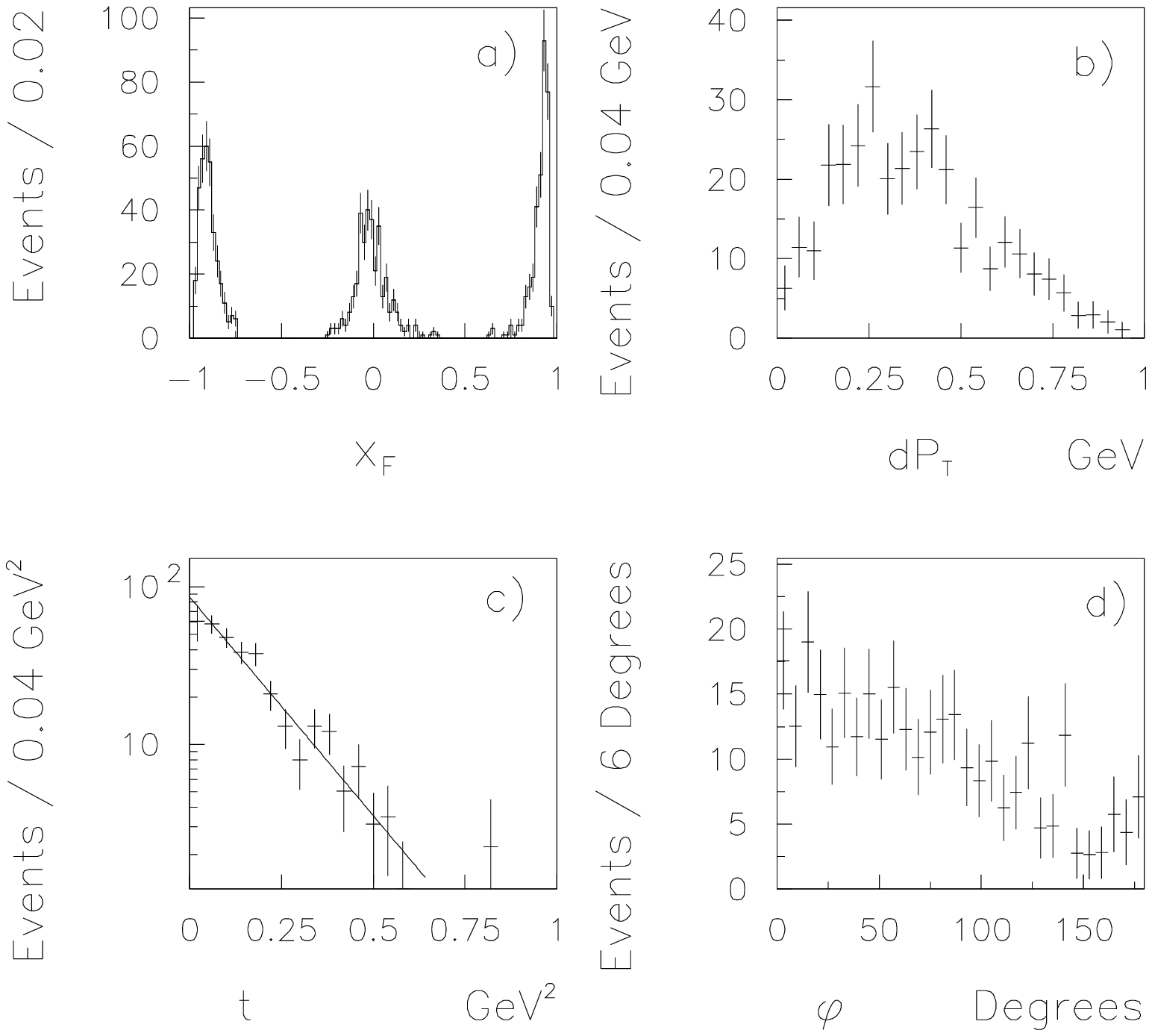,height=22cm,width=17cm}
\end{center}
\begin{center} {Figure 4} \end{center}
\end{document}